# Rebound indentation problem for a viscoelastic half-space and axisymmetric indenter — Solution by the method of dimensionality reduction

**I.I. Argatov and V.L. Popov**[*]

*Institut für Mechanik, Technische Universität Berlin, Straße des 17. Juni 135, 10623 Berlin, Germany*

The method of dimensionality reduction (MDR) is extended for the axisymmetric frictionless unilateral Hertz-type contact problem for a viscoelastic half-space and an arbitrary axisymmetric rigid indenter under the assumption that an arbitrarily evolving in time circular contact area remains singly connected during the whole process of indentation. In particular, the MDR is applied to study in detail the so-called rebound indentation problem, where the contact radius has a single maximum. It is shown that the obtained closed-form analytical solution for the rebound indentation displacement (recorded in the recovery phase, when the contact force vanishes) does not depend on the indenter shape.

## 1 Introduction

Contact problems for a viscoelastic half-space and an axisymmetric indenter have been used in various applications, including the indentation testing of time-dependent materials [1], [2] and the grasping contact analysis of biological tissues [3]. Since the contact problem formulation with an *a priori* unknown contact area includes consideration of the contact radius history, solving such a problem turns out to be a nontrivial task even in the case of a spherical indenter [4], [5].

In recent years, the method of dimensionality reduction (MDR) has been developed by Popov and Heß [6], [7] for effective dealing with axisymmetric unilateral Hertz-type elastic contact problems. This method maps a given contact problem into some mathematical model of 1D contact for a generalized discrete linear elastic foundation and thereby reducing the original problem complexity to that of the Winkler elastic foundation contact model. The MDR has been rigorously established [7], [8] in the elastic case, whereas the viscoelastic case requires a special consideration and justification. In the case of frictionless unilateral viscoelastic contact for a spherical indenter, the MDR mapping rule of Popov was formulated for monotonic loadings (when the contact radius is not decreasing in time), based on the Lee–Radok elastic-viscoelastic correspondence principle [9].

The purpose of writing this paper is twofold. The first is to extend the MDR formalism for non-monotonic loadings. In viscoelastic contact problems, the loading protocol plays an important role and, generally speaking, the Lee–Radok elastic-viscoelastic correspondence principle is not applicable in situations when the loading results in the reduction of the contact area. The second purpose is to give a solution of the rebound indentation problem, which models the so-called rebound indentation test [10] with a jump-like reduction of the contact area after some constant rate indentation. Recently, the rebound indentation problem was studied in the cases of cylindrical [11] and spherical [12] indenters.

---

[*] Corresponding author  E-mail: v.popov@tu-berlin.de, Phone: +49 30 314 21480,  Fax: +49 30 314 72575

Here we generalize the obtained solution [12] for the case of arbitrary axisymmetric indenter, which produces a circular contact area during the whole process of indentation.

## 2 Formulation of the unilateral axisymmetric contact problem for a viscoelastic half-space

We consider an isotropic linear viscoelastic half-space indented by a frictionless rigid indenter with the axially symmetric shape function $f(r)$, $r \in [0, +\infty)$, such that $f(0) = 0$ and $f'(r) > 0$ for $r > 0$. Under these assumptions, the contact area, $\omega(t)$, will be circular with *a priori* unknown radius $a(t)$.

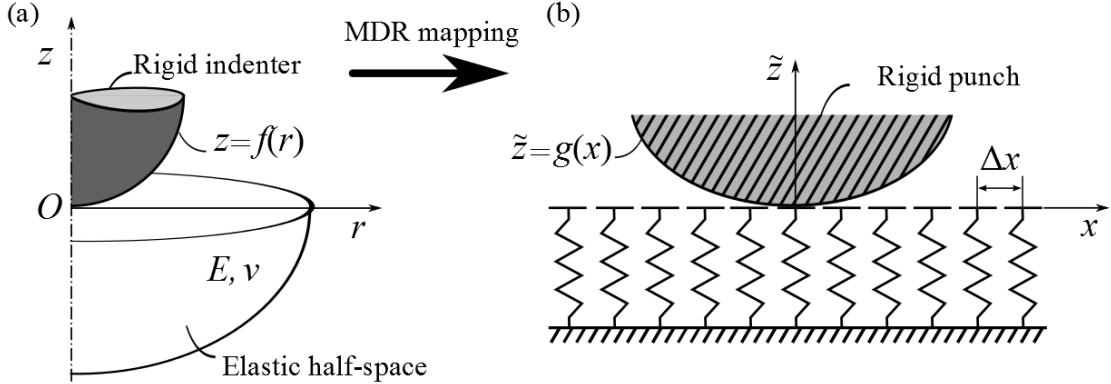

**Fig. 1** Initial contact configuration: (a) Axisymmetric contact problem; (b) MDR-based equivalent 1D contact model.

For the sake of simplicity, we assume that Poisson's ratio, $v$, of the half-space material is time independent. Then, by applying the elastic-viscoelastic correspondence principle [13], the vertical displacement of the half-space surface, $U_z(r,t)$, can be expressed through the contact pressure, $p(r,t)$, as follows:

$$U_z(r,t) = \frac{1}{\pi E_0^*} \int_{0^-}^{t} \phi(t-\tau) \frac{\partial}{\partial \tau} \iint_{\omega(\tau)} \frac{p(\rho,t)}{R} dS\, d\tau. \tag{1}$$

Here, $E_0^* = E_0 / (1-v^2)$ is the instantaneous effective elastic modulus, $E_0$ is the instantaneous elastic modulus, $R$ is the distance between the point of observation (with the coordinate $r$) and the point of integration (with the radial coordinate $\rho$), $dS = \rho\, d\rho\, d\varphi$ is the area element in polar coordinates $(\rho, \varphi)$, and $\phi(t)$ is the normalized creep function. The lower limit $0^-$ is used in the integration in (1) to account for possible jump in the contact pressure at $t = 0$, while it is assumed that for $t < 0$, the viscoelastic half-space is stress-free.

Let $E(t)$ denote the relaxation modulus such that $E(t) = E_0 \psi(t)$, where $\psi(t)$ is the normalized relaxation function. By definition, the normalized creep function $\phi(t)$ is the reciprocal of $\psi(t)$ so that if $v(0^-) = 0$ and

$$u(t) = \int_{0^-}^{t} \phi(t-\tau) \frac{\partial v}{\partial \tau}(\tau) \, d\tau, \tag{2}$$

then

$$v(t) = \int_{0^-}^{t} \psi(t-\tau) \frac{\partial u}{\partial \tau}(\tau) \, d\tau. \tag{3}$$

When the indenter is pressed into the half-space by an amount, $w(t)$, the contact problem then is to find $p(r,t)$ such that $p(r,t) > 0$ and $U_z(r,t) = w(t) - f(t)H(t)$ inside the contact area $\omega(t)$, while $p(r,t) = 0$ outside the contact area and the free half-space surface does not penetrate the surface of the indenter. Here, $H(t)$ is the Heaviside step-function, which is 0 for $t < 0$ and 1 for $t \geq 0$. Therefore, the contact pressure density should satisfy the following conditions of unilateral contact:

$$p(r,t) \geq 0, \quad 0 \leq r < +\infty, \tag{4}$$

$$p(r,t) > 0 \implies \frac{1}{\pi E_0^*} \int_{0^-}^{t} \phi(t-\tau) \frac{\partial}{\partial \tau} \iint_{\omega(\tau)} \frac{p(\rho,\tau)}{R} \, dS \, d\tau = w(t) - f(r)H(t), \tag{5}$$

$$p(r,t) = 0 \implies \frac{1}{\pi E_0^*} \int_{0^-}^{t} \phi(t-\tau) \frac{\partial}{\partial \tau} \iint_{\omega(\tau)} \frac{p(\rho,\tau)}{R} \, dS \, d\tau \geq w(t) - f(r)H(t). \tag{6}$$

It should be also noted [14] that $w(t)$ must involve the Heaviside function factor $H(t)$ so that the right-hand sides of (5) and (6) vanish for $t < 0$ on the whole half-space surface. In other words, we will assume that $w(t) = 0$ for $t < 0$.

The indenter displacement $w(t)$ should be determined from the equilibrium equation

$$\iint_{\omega(t)} \frac{p(\rho,t)}{R} \, dS = F(t), \tag{7}$$

where $F(t)$ is the total force applied on the indenter, such that $F(t) = 0$ for $t < 0$.

Thus, the unilateral contact problem consists of finding the histories for the contact pressure $p(r,t)$, $r \in [0, +\infty)$, $t \in [0, +\infty)$, the radius $a(t)$ of the contact area $\omega(t)$ (where the contact pressure is positive), and the indenter displacement $w(t)$, which for the known contact force history $F(t)$, $t \in [0, +\infty)$, and the prescribed indenter shape function $f(r)$ satisfy the relations (4)–(7).

## 3 MDR mapping to the 1D contact problem for a viscoelastic Winkler foundation

First, following [7], the indenter shape function $f(r)$, $r \in [0, +\infty)$, is replaced by an equivalent 1D profile $g(x)$, $x \in (-\infty, +\infty)$, for the equivalent rigid punch according to the Popov–Geike–Heß mapping rule

$$g(x) = |x| \int_0^{|x|} \frac{f'(r) \, dr}{\sqrt{x^2 - r^2}}. \tag{8}$$

Second, we introduce the so-called linear viscoelastic Winkler foundation consisting of independent identical viscoelastic springs that are fixed to a rigid substrate and are separated from one another by a small distance $\Delta x$ (called the discretization step). The relaxation stiffness of every individual spring element is given by

$$\Delta k_z(t) = E_0^* \psi(t) \Delta x, \tag{9}$$

where $E_0^*$ and $\psi(t)$ are the instantaneous elastic modulus and the normalized relaxation function of the viscoelastic half-space.

In this way, the reaction force, $\Delta f_N(x,t)$, of the individual spring element with a coordinate $x$ is related to the spring contraction, $u_z(x,t)$, as follows:

$$\Delta f_N(x,t) = \int_{0^-}^{t} \Delta k_z(t-\tau) \frac{\partial u_z}{\partial \tau}(x,\tau) \, d\tau. \tag{10}$$

Moreover, let us introduce the distributed foundation reaction

$$q(x,t) = \frac{\Delta f_N(x,t)}{\Delta x}. \tag{11}$$

In view of (9) and (11), Eq.(10) can be represented in the form

$$q(x,t) = E_0^* \int_{0^-}^{t} \psi(t-\tau) \frac{\partial u_z}{\partial \tau}(x,\tau) \, d\tau. \tag{12}$$

When the equivalent rigid punch with the profile function $g(x)$ is pressed into the viscoelastic Winkler foundation, its local indentation is given by

$$u_z(x,t) = \frac{1}{E_0^*} \int_{0^-}^{t} \phi(t-\tau) \frac{\partial q}{\partial \tau}(x,\tau) \, d\tau, \tag{13}$$

where $\phi(t)$ is the corresponding normalized creep function.

Since, in unilateral contact, the spring reaction forces cannot take negative values, we have (cf. formula (4))

$$q(x,t) \geq 0, \quad x \in (-\infty, +\infty). \tag{14}$$

By analogy with the unilateral contact conditions (5) and (6), we write out the relations

$$q(x,t) > 0 \implies \frac{1}{E_0^*} \int_{0^-}^{t} \phi(t-\tau) \frac{\partial q}{\partial \tau}(x,\tau) \, d\tau = w(t) - g(x) H(t), \tag{15}$$

$$q(x,t) = 0 \implies \frac{1}{E_0^*} \int_{0^-}^{t} \phi(t-\tau) \frac{\partial q}{\partial \tau}(x,\tau) \, d\tau \geq w(t) - g(x) H(t). \tag{16}$$

Inside the contact interval $D(t) = (-a(t), a(t))$, we have

$$q(x,t) > 0, \quad x \in (-a(t), a(t)), \tag{17}$$

while $q(x,t) = 0$ outside the contact interval.

The total normal force needed to press the equivalent punch against the viscoelastic Winkler foundation, thereby achieving the indentation depth $w(t)$, is evaluated as the sum of all contributions of single springs $\Delta f_N(x,t)$ for $x \in (-a(t), a(t))$, which as $\Delta x \to 0$, reduces to the integral

$$F(t) = \int_{-a(t)}^{a(t)} q(x,t) \, dx. \tag{18}$$

Thus, the unilateral equivalent 1D contact problem consists of finding the contact reaction $q(x,t)$, $x \in (-\infty, +\infty)$, the contact interval history $(-a(t), a(t))$, and the equivalent punch displacement $w(t)$, which for the known contact force history $F(t)$, $t \in [0, +\infty)$, and the prescribed punch profile function $g(x)$ satisfy the relations (14)–(18).

It should be emphasized that the MDR establishes direct equivalence relations between the half-length of the 1D contact interval and the radius of the original contact area (that is why, both characteristics are denoted by the same symbol $a(t)$) as well as between the normal contact forces denoted by $F(t)$. In other words, the contact force, $F(t)$, the indentation displacement, $w(t)$, and the characteristic size of the contact zone, $a(t)$, in both contact problems take the same values.

Finally, the contact pressure density $p(r,t)$ is expressed in terms of the 1D viscoelastic foundation contact reaction $q(x,t)$ via the transformation

$$p(r,t) = -\frac{1}{\pi} \int_{r}^{\infty} \frac{1}{\sqrt{x^2 - r^2}} \frac{\partial q}{\partial x}(x,t) \, dx. \tag{19}$$

We note that the mentioned above equivalence has been rigorously established [7] in the base-case problem of axisymmetric normal frictionless elastic Hertz-type contact. In the present paper, we prove this equivalence for the viscoelastic case.

**4 Solution of the equivalent 1D contact problem in the case of monotonically increasing contact area**

In light of (15) and (17), the governing integral equation takes the form

$$\frac{1}{E_0^*} \int_{0^-}^{t} \phi(t-\tau) \frac{\partial q}{\partial \tau}(x,\tau) \, d\tau = w(t) - g(x)H(t), \qquad |x| \leq a(t). \tag{20}$$

In view of the non-penetration condition (16) and the fact that outside the contact interval, the contact reaction vanishes, we can extend Eq.(20) to the whole surface of the viscoelastic foundation, i.e., for $x \in (-\infty, +\infty)$, as follows:

$$\frac{1}{E_0^*} \int_{0^-}^{t} \phi(t-\tau) \frac{\partial q}{\partial \tau}(x,\tau) \, d\tau = \left( w(t) - g(x)H(t) \right)_+. \tag{21}$$

Here, $(s)_+ = (s + |s|)/2$ is the positive part function.

Assuming that $g(x)$, $x \in [0, +\infty)$, is a strictly increasing function, we conclude that the half-length of the contact interval $a(t)$ is related to the punch displacement $w(t)$ by the formula

$$g(a(t)) = w(t), \tag{22}$$

or (see formula (8))

$$a(t) \int_0^{a(t)} \frac{f'(r) \, dr}{\sqrt{a(t)^2 - r^2}} = w(t), \tag{23}$$

where $f(r)$ is the original indenter's shape function.

Now, as the right-hand side of the integral equation (20) makes sense for all $x \in (-\infty, +\infty)$, we can write out its solution using Eqs.(2) and (3) as follows:

$$q(x,t) = E_0^* \int_{0^-}^{t} \psi(t-\tau) \frac{\partial}{\partial \tau} \left( w(\tau) - g(x)H(\tau) \right)_+ d\tau. \tag{24}$$

The substitution of (24) (with (22) taken into account) into Eq.(18) yields the contact force in the form

$$F(t) = 2E_0^* \int_{0^-}^{t} \psi(t-\tau) \frac{\partial}{\partial \tau} \left\{ a(\tau) g(a(\tau)) - \int_{0}^{a(\tau)} g(x) \, dx \right\} d\tau. \tag{25}$$

Now, substituting here the expression for $g(x)$ provided by the mapping rule (8), we eventually arrive at the formula

$$F(t) = 2E_0^* \int_{0^-}^{t} \psi(t-\tau) \frac{\partial}{\partial \tau} \int_{0}^{a(\tau)} \frac{f'(r) r^2}{\sqrt{a(\tau)^2 - r^2}} \, dr \, d\tau. \tag{26}$$

Further, differentiating the foundation reaction, we readily get

$$\frac{\partial q}{\partial x}(x,t) = -E_0^* \int_{0^-}^{t} \psi(t-\tau) \frac{\partial}{\partial \tau} \{ g'(x) H(\tau) H(a(\tau) - x) \} d\tau, \tag{27}$$

and the substitution of the obtained integral into Eq.(19) results in the formula

$$p(r,t) = \frac{E_0^*}{\pi} \int_{0^-}^{t} \psi(t-\tau) \frac{\partial}{\partial \tau} \int_{r}^{a(\tau)} \frac{g'(x)}{\sqrt{x^2 - r^2}} \, dx \, d\tau. \tag{28}$$

Recall that $a(t)$ is assumed to involve the Heaviside step function factor so that $a(t) = 0$ for $t < 0$ as well as $F(t) = 0$ for $t < 0$. We also note that it is not hard to see that in the case of a spherical indenter, the derived result coincides with the solution originally obtained by Lee and Radok [9].

Following Ting [14], in view of Eqs.(23), (24), (26), and (28), we introduce the auxiliary notation

$$w_e(t) = a(t) \int_{0}^{a(t)} \frac{f'(r) \, dr}{\sqrt{a(t)^2 - r^2}}, \tag{29}$$

$$F_e(t) = 2E_0^* \int_{0}^{a(t)} \frac{f'(r) r^2}{\sqrt{a(t)^2 - r^2}} \, dr, \tag{30}$$

$$q_e(x,t) = E_0^* \left( g(a(t)) - g(x) H(t) \right)_+, \tag{31}$$

$$p_e(r,t) = -\frac{1}{\pi} \int_{r}^{a(t)} \frac{1}{\sqrt{x^2 - r^2}} \frac{\partial q_e}{\partial x}(x,t) \, dx. \tag{32}$$

Here, a subscript $e$ distinguishes elastic solutions from viscoelastic solutions.

Thus, in light of (29)–(32), the solution (22), (26), and (28), which was obtained using the MDR, can be recast in the form

$$w(t) = w_e(t), \tag{33}$$

$$F(t) = \int_{0^-}^{t} \psi(t-\tau) \frac{\partial F_e}{\partial \tau}(\tau) \, d\tau \, , \tag{34}$$

$$p(r,t) = \int_{0^-}^{t} \psi(t-\tau) \frac{\partial p_e}{\partial \tau}(r,\tau) \, d\tau \, . \tag{35}$$

To this end, apart from the notation, the solution (33)–(35) recovers the solution obtained by Ting [14]. That is to say that the MDR correctly solves the contact problem under consideration in the case of monotonic loading.

## 5 Solution of the equivalent 1D contact problem when the contact radius has a single maximum

The solution constructed in the previous section is valid for $0 \le t < t_m$, where $t_m$ is the time at which $a(t)$ is a maximum (see Fig.2). Now, we consider the contact problem on the next interval $t_m \le t \le t_n$, where $t_n$ is the time at which $a(t)$ becomes a minimum. In this interval, the half-length of the 1D contact interval $a(t)$ (i.e., the contact radius of the original circular contact area) monotonically decreases. Following [14], the governing integral equation (20) is represented in the form

$$E_0^* \int_{0^-}^{t} \psi(t-\tau) \frac{\partial u_z}{\partial \tau}(x,\tau) \, d\tau = q(x,t), \qquad |x| \le a(t), \tag{36}$$

where

$$u_z(x,t) = w(t) - g(x) H(t) \, . \tag{37}$$

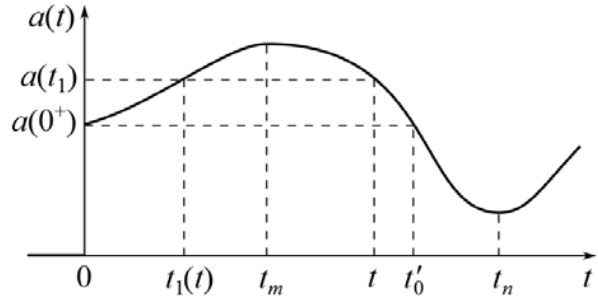

**Fig. 2** A stepwise non-monotonic history of the contact radius evolution.

The key idea [4] is to rewrite Eq.(36) as follows:

$$\int_{0^-}^{t_1(t)} \psi(t-\tau) \frac{\partial u_z}{\partial \tau}(x,\tau) \, d\tau + \int_{t_1(t)}^{t} \psi(t-\tau) \frac{\partial u_z}{\partial \tau}(x,\tau) \, d\tau = \frac{1}{E_0^*} q(x,t) \, . \tag{38}$$

Here, $t_1(t)$ is the solution of the equation

$$a(t_1) = a(t), \qquad t > t_m, \qquad t_1 < t_m. \tag{39}$$

Substituting the expression (37) into the second term on the left-hand side of Eq.(38), we get

$$\int_{0^-}^{t_1(t)} \psi(t-\tau)\frac{\partial u_z}{\partial \tau}(x,\tau)\,d\tau + \int_{t_1(t)}^{t} \psi(t-\tau)\frac{\partial w}{\partial \tau}(\tau)\,d\tau - g(x)\int_{t_1(t)}^{t} \psi(t-\tau)\frac{\partial H}{\partial \tau}(\tau)\,d\tau = \frac{1}{E_0^*}q(x,t). \tag{40}$$

The last term on the left-hand side of Eq.(40) depends on whether $t < t_0'$ or $t > t_0'$ (see Fig.2). Namely, it is zero if $t_1(t) > 0$ and is $g(x)\psi(t)$ if $t_1(t) = 0^-$. Thus, in a way similar to [14], we arrive at the equation

$$\int_{t_1(t)}^{t} \psi(t-\tau)\frac{\partial w}{\partial \tau}(\tau)\,d\tau - g(x)\psi(t)H(t-t_0') = \frac{1}{E_0^*}q(x,t) - \int_{0^-}^{t_1(t)} \psi(t-\tau)\frac{\partial}{\partial \tau}u_z(x,\tau)\,d\tau, \qquad |x| \leq a(t). \tag{41}$$

Since $t_1(t) < t_m$, the local indentation $u_z(x,\tau)$ is known on the whole interval of integration $\tau \in [0^-, t_1(t))$ on the right-hand side of the above equation, and, in view of (22) and (31), it can be represented as

$$u_z(x,\tau) = \frac{1}{E_0^*}q_e(x,\tau), \qquad |x| \leq a(t), \qquad 0^- < \tau < t_1(t), \tag{42}$$

while

$$q_e(x,\tau) = 0, \qquad a(\tau) \leq x \leq a(t). \tag{43}$$

Let us first consider the case $t_m < t < t_0'$, when in view of (42), Eq.(41) reduces to

$$E_0^* \int_{t_1(t)}^{t} \psi(t-\tau)\frac{\partial w}{\partial \tau}(\tau)\,d\tau = q(x,t) - \int_{0^-}^{t_1(t)} \psi(t-\tau)\frac{\partial q_e}{\partial \tau}(x,\tau)\,d\tau, \qquad |x| \leq a(t), \tag{44}$$

where the left-hand side does not depend on $x$.

Substituting the value $x = a(t)$ into Eq.(44) and taking into account (43) and the boundary condition $q(a(t),t) = 0$, we obtain

$$\int_{t_1(t)}^{t} \psi(t-\tau)\frac{\partial w}{\partial \tau}(\tau)\,d\tau = 0, \tag{45}$$

and correspondingly

$$q(x,t) = \int_{0^-}^{t_1(t)} \psi(t-\tau)\frac{\partial q_e}{\partial \tau}(x,\tau)\,d\tau. \tag{46}$$

Using Eq.(33), which is valid up to the time $t_m$, we rewrite Eq.(45) as

$$\int_{t_m}^{t} \psi(t-\tau)\frac{\partial w}{\partial \tau}(\tau)\,d\tau = -\int_{t_1(t)}^{t_m} \psi(t-\tau)\frac{\partial w_e}{\partial \tau}(\tau)\,d\tau, \tag{47}$$

and following the procedure described in detail by Ting [14], we finally arrive at the equation

$$w(t) = w_e(t) - \int_{t_m}^{t} \phi(t-\tau)\frac{\partial}{\partial \tau}\int_{t_1(\tau)}^{\tau} \psi(t-\eta)\frac{\partial w_e}{\partial \eta}(\eta)\,d\eta\,d\tau. \tag{48}$$

Substituting the foundation reaction (46) into the inverse mapping formula (19) and taking into account relations (27), (31), and (32), we obtain the contact pressure density

$$p(r,t) = \int_{0^-}^{t_1(t)} \psi(t-\tau)\frac{\partial p_e}{\partial \tau}(r,\tau)\,d\tau, \tag{49}$$

and, correspondingly, the contact force is given by

$$F(t) = \int_{0^-}^{t_1(t)} \psi(t-\tau)\frac{\partial F_e}{\partial \tau}(\tau)\,d\tau, \tag{50}$$

where $p_e(r,t)$ and $F_e(t)$ are defined by (32) and (30), respectively.

Now, let us consider the case $t_0' < t < t_n$, when in view of the convention $t_1(t) = 0^-$ for $t_0' < t < t_n$, Eq.(41) takes the form

$$\int_{0^-}^{t} \psi(t-\tau)\frac{\partial w}{\partial \tau}(\tau)\,d\tau - g(x)\psi(t) = \frac{1}{E_0^*}q(x,t), \quad |x| \leq a(t). \tag{51}$$

Substituting the value $x = a(t)$ into Eq.(51), we readily see that its right-hand side vanishes, and we obtain

$$\int_{0^-}^{t} \psi(t-\tau)\frac{\partial w}{\partial \tau}(\tau)\,d\tau = g(a(t))\psi(t), \tag{52}$$

or, taking into account the notation (29), we get

$$\int_{t_m}^{t} \psi(t-\tau)\frac{\partial w}{\partial \tau}(\tau)\,d\tau = -\int_{0^-}^{t_m} \psi(t-\tau)\frac{\partial w_e}{\partial \tau}(\tau)\,d\tau + w_e(t)\psi(t).$$

Comparing the above equation with Eq.(47), we see that to cover both cases $t_m < t < t'_0$ and $t'_0 < t < t_n$, we rewrite them as follows:

$$\int_{t_m}^{t} \psi(t-\tau) \frac{\partial w}{\partial \tau}(\tau) \, d\tau = - \int_{t_1(t)}^{t_m} \psi(t-\tau) \frac{\partial w_e}{\partial \tau}(\tau) \, d\tau + w_e(t) \psi(t) H(t-t'). \tag{53}$$

Apart from the notation, Eq.(53) coincides with the intermediate equation derived by Ting [14], who further transformed it into another form, which is more suitable for numerical solution.

Returning to Eq.(51) and taking into account Eq.(52), we arrive at the equation

$$q(x,t) = E_0^* \psi(t) \bigl( g(a(t)) - g(x) \bigr)_+, \tag{54}$$

which in view of the notation (31) can be represented as

$$q(x,t) = \psi(t) q_e(x,t). \tag{55}$$

Finally, substituting the foundation reaction (55) into the inverse mapping formula (19) and taking into account relations (27), (31), and (32), we readily get the solution

$$p(r,t) = \psi(t) p_e(r,t), \tag{56}$$

which completely agrees with the solution originally found by Ting [14], meaning that the MDR correctly solves the contact problem on the sage of unloading.

### 6 Solution of the rebound indentation problem

Following Brown *et al.* [10], we consider the rebound indentation test, which is composed of two stages (see Fig.3). In the first stage, which is called the indentation phase, the viscoelastic half-space is indented at a constant rate, $v_0$, to a maximum indentation depth, $w_m$. In other words, the indenter displacement is assumed to follow the linear law

$$w^{(1)}(t) = v_0 t, \qquad 0 \le t < t_m, \tag{57}$$

while the maximum indenter displacement, $w_m = w^{(1)}(t_m)$, is given by

$$w_m = v_0 t_m. \tag{58}$$

At the end of the first stage, the load is assumed to be immediately removed, and the indenter displacement is recorded during the second stage, called the recovery phase, when we have

$$F^{(2)}(t) = 0, \qquad t > t_m. \tag{59}$$

In what follows we make use of the notation adopted in refs. [11], [12].

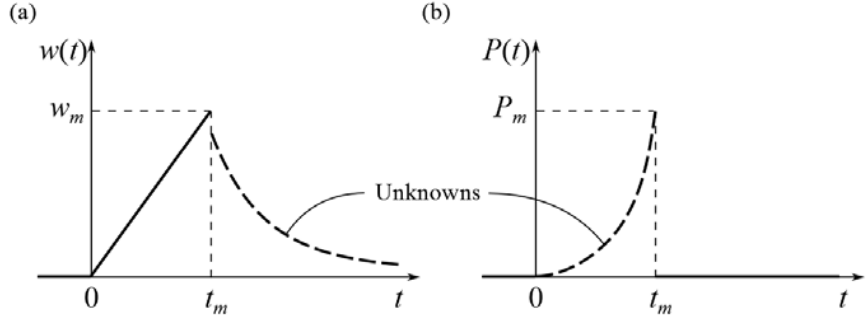

**Fig. 3** Schematics of the rebound indentation test.

By formula (34), the contact force during the indentation phase changes according to the law

$$F^{(1)}(t) = \int_{0^-}^{t} \psi(t-\tau)\frac{\partial F_e}{\partial \tau}(\tau)\,d\tau, \qquad 0 \le t < t_m, \tag{60}$$

where the elastic solution $F_e(t)$, corresponding to the contact radius $a(t)$, is given by (30), while the contact radius shows itself as a solution to the equation

$$w_e(t) = w^{(1)}(t), \tag{61}$$

where $w^{(1)}(t)$ and $w_e(t)$ are provided by (57) and (29), respectively.

The substitution of (30) and (29), (57) into Eqs.(60) and (61), respectively, yields

$$F^{(1)}(t) = 2E_0^* \int_{0^-}^{t} \psi(t-\tau)\frac{\partial}{\partial \tau}\int_0^{a(\tau)} \frac{f'(r)r^2}{\sqrt{a(\tau)^2 - r^2}}\,dr\,d\tau, \tag{62}$$

$$a(t)\int_0^{a(t)} \frac{f'(r)\,dr}{\sqrt{a(t)^2 - r^2}} = v_0 t, \qquad 0 \le t < t_m. \tag{63}$$

Now, let us assume that the unloading of the punch is achieved in a finite time interval $t_m \le t \le t_n$, at the end of which the contact force $F^{(2)}(t)$ reaches the zero value, i.e., $F^{(2)}(t_n) = 0$. Since $a(0^+) = 0$ due to a gradually increasing indentation loading protocol (57), in the unloading stage $t_m \le t \le t_n$, we may make use of the solution (48) and (50), that is

$$w^{(2)}(t) = w_e(t) - \int_{t_m}^{t} \phi(t-\tau)\frac{\partial}{\partial \tau}\int_{t_1(\tau)}^{\tau} \psi(t-\eta)\frac{\partial w_e}{\partial \eta}(\eta)\,d\eta\,d\tau, \tag{64}$$

$$F^{(2)}(t) = \int_{0^-}^{t_1(t)} \psi(t-\tau)\frac{\partial F_e}{\partial \tau}(\tau)\,d\tau, \qquad t_m \le t \le t_n, \tag{65}$$

where $t_1(t)$ solves the equation

$$a^{(1)}(t_1) = a^{(2)}(t), \qquad t_1 \leq t_m \leq t. \tag{66}$$

In light of (29) and (66), we have

$$w_e(t) = a^{(2)}(t) \int_0^{a^{(2)}(t)} \frac{f'(r)\,\mathrm{d}r}{\sqrt{a^{(2)}(t)^2 - r^2}} = a^{(1)}(t_1) \int_0^{a^{(1)}(t_1)} \frac{f'(r)\,\mathrm{d}r}{\sqrt{a^{(1)}(t_1)^2 - r^2}}$$

$$= w_e^{(1)}(t_1), \qquad t_m \leq t \leq t_n.$$

Correspondingly, Eq.(64) can be rewritten as

$$w^{(2)}(t) = w_e^{(1)}(t_1(t)) - \int_{t_m}^{t} \phi(t-\tau) \frac{\partial}{\partial \tau} \left\{ \int_{t_1(\tau)}^{t_m} \psi(t-\eta) \frac{\partial w_e^{(1)}}{\partial \eta}(\eta)\,\mathrm{d}\eta + \int_{t_m}^{\tau} \psi(t-\eta) \frac{\partial w_e^{(1)}}{\partial \eta}(t_1(\tau))\,\mathrm{d}\eta \right\} \mathrm{d}\tau.$$

Thus, making use of the method developed in [12], we further transform this relation to the form

$$w^{(2)}(t) = w^{(1)}(t_1(t)) + \int_{t_1(t)}^{t_m} G(t-\eta, \tau(\eta) - \eta) \frac{\mathrm{d}w^{(1)}}{\mathrm{d}\eta}(\eta)\,\mathrm{d}\eta,$$

where $G(t,\tau)$ is Greenwood's function [15] given by

$$G(t,\tau) = 1 - \phi(0)\psi(t) + \int_\tau^t \psi(\xi) \frac{\partial \phi}{\partial \xi}(t-\xi)\,\mathrm{d}\xi. \tag{67}$$

In the rebound indentation test, under the assumption of instantaneous unloading, we arrive at the following result [12]:

$$w^{(2)}(t) = \int_0^{t_m} G(t-\eta, t_m - \eta) \frac{\mathrm{d}w^{(1)}}{\mathrm{d}\eta}(\eta)\,\mathrm{d}\eta.$$

For the constant rate loading indentation phase (58), the above formula simplifies to

$$w^{(2)}(t) = v_0 \int_0^{t_m} G(t-\eta, t_m - \eta)\,\mathrm{d}\eta. \tag{68}$$

It is interesting to observe that formula (68) coincides with the solution for the rebound displacement obtained in [12] in the case of a spherical punch. In fact, formula (68) does not depend on the indenter shape function. This, in particular, means that formula (68) must also hold for a cylindrical indenter, which can be regarded as the limiting situation for indenters of power-law shape. However, in the latter case, the following solution was obtained by Argatov and Mishuris [11]:

$$w^{(2)}(t) = v_0 \int_0^{t_m} [\phi(t-\tau) - \phi(t-t_m)] \psi(\tau) \, d\tau . \tag{69}$$

Let us show that formulas (68) and (69) are equivalent. Indeed, in light of (67), we have

$$G(t-\eta, t_m - \eta) = 1 - \phi(0)\psi(t-\eta) + \int_{t_m-\eta}^{t-\eta} \psi(\xi) \frac{\partial \phi}{\partial \xi}(t-\eta-\xi) \, d\xi$$

$$= 1 - \phi(0)\psi(t-\eta) + \int_{t_m-\eta}^{t-\eta} \psi(\xi) \frac{\partial \phi}{\partial \eta}(t-\eta-\xi) \, d\xi .$$

Thus, the integral in (68) can be represented as

$$\int_0^{t_m} G(t-\eta, t_m - \eta) \, d\eta = t_m - \phi(0) \int_0^{t_m} \psi(t-\eta) \, d\eta + I_1 - I_2 , \tag{70}$$

where we introduced the notation

$$I_1 = \int_0^{t_m} d\eta \int_0^{t-\eta} \psi(\xi) \frac{\partial \phi}{\partial \eta}(t-\eta-\xi) \, d\xi ,$$

$$I_2 = \int_0^{t_m} d\eta \int_0^{t_m-\eta} \psi(\xi) \frac{\partial \phi}{\partial \eta}(t-\eta-\xi) \, d\xi .$$

By changing the integration variables, we obtain

$$I_1 = \int_t^{t-t_m} dy \int_0^y \psi(\xi) \frac{\partial}{\partial y} \phi(y-\xi) \, d\xi ,$$

$$I_2 = -\int_0^{t_m} dy \int_0^y \psi(\xi) \frac{\partial}{\partial y} \phi(t-t_m+y-\xi) \, d\xi .$$

Now, making use of the formula

$$\int_0^\tau \frac{\partial f}{\partial \tau}(\tau, \xi) \, d\xi = -f(\tau, \tau) + \frac{d}{d\tau} \int_0^\tau f(\tau, \xi) \, d\xi ,$$

we convert the integrals $I_1$ and $I_2$ to the forms

$$I_1 = \int_0^{t-t_m} \psi(\xi) \phi(t-t_m-\xi) \, d\xi - \int_0^t \psi(\xi) \phi(t-\xi) \, d\xi - \phi(0) \int_t^{t-t_m} \psi(\eta) \, d\eta ,$$

$$I_2 = -\int_0^{t_m} \psi(\xi)\phi(t-\xi)\,\mathrm{d}\xi + \int_0^{t_m} \psi(\eta)\phi(t-t_m)\,\mathrm{d}\eta.$$

Finally, substituting the obtained expressions for $I_1$ and $I_2$ into formula (70) and using the known relation

$$\int_0^t \phi(t-\tau)\psi(\tau)\,\mathrm{d}\tau = t,$$

we arrive at formula (69). Thus, the two forms (68) and (69) of the solution are equivalent.

### 7 Justification of the MDR in the case of axisymmetric Hertz-type viscoelastic contact

Let us return to the 1D contact problem for the viscoelastic foundation (14)–(16). Due to the symmetry assumption (see, in particular, formula (8)) and the monotonicity assumption about the shape function $f(r)$, the 1D contact zone will constitute a continuous interval, which is denoted by $D(t) = (-a(t), a(t))$.

Now, we introduce a singular kernel

$$K(x,\xi) = \delta(\xi - x), \tag{71}$$

where $\delta(s)$ is the Dirac delta function such that

$$\int_{-\infty}^{+\infty} \delta(\xi - x)v(\xi)\,\mathrm{d}\xi = v(x)$$

for every compactly supported continuous function $v(x)$.

Then, Eq.(13) can be represented as

$$u_z(x,t) = \frac{1}{E_0^*}\int_{0^-}^{t} \phi(t-\tau)\frac{\partial}{\partial \tau}\int_{D(\tau)} K(x,\xi)q(\xi,\tau)\,\mathrm{d}\xi\,\mathrm{d}\tau, \tag{72}$$

whereas Eq.(36) takes the form

$$E_0^*\int_{0^-}^{t} \psi(t-\tau)\frac{\partial u_z}{\partial \tau}(x,\tau)\,\mathrm{d}\tau = \int_{D(t)} K(x,\xi)q(\xi,t)\,\mathrm{d}\xi. \tag{73}$$

In light of (72) and (73), we can apply the algorithm developed by Ting [16] to express the sought-for solution of the 1D viscoelastic contact problem in terms of the elastic solution $w_e(t)$, $F_e(t)$, and $q_e(x,t)$, which satisfies the relations

$$u_z^e(x,t) = \frac{1}{E_0^*} \int_{D(t)} K(x,\xi) q_e(\xi,t) \, d\xi,$$

$$q_e(x,t) > 0, \quad x \in D(t),$$

$$u_z^e(x,t) \begin{cases} = w_e(t) - g(x)H(t), & x \in D(t), \\ \geq w_e(t) - g(x)H(t), & x \notin D(t). \end{cases}$$

In the Winkler foundation case (71), we evidently have

$$q_e(x,t) = E_0^* \left( g(a(t)) - g(x)H(t) \right)_+,$$

$$w_e(t) = g(a(t)),$$

$$F_e(t) = \int_{-a(t)}^{a(t)} q_e(x,t) \, dx.$$

Now, taking into account the definition of the function $g(x)$ (see the mapping rule (8)) and the inverse-mapping relation (see formulas (19) and (32))

$$p_e(r,t) = -\frac{1}{\pi} \int_r^{a(t)} \frac{1}{\sqrt{x^2 - r^2}} \frac{\partial q_e}{\partial x}(x,t) \, dx,$$

as well as the fact that the pair of functions $q_e(x,t)$ and $q(x,t)$, $p_e(r,t)$ and $p(r,t)$ are related by the same combinations of the Boltzmann hereditary integral operators, we conclude that the MDR (based on Ting's solution algorithm) yields the solution to the original viscoelastic contact problem (4)–(7).

**8 Discussion**

First of all, we note that the assumption of constant Poisson's ratio can be relaxed. This assumption is often made for elastomers, which usually can be considered to be incompressible materials, so that $\nu = 0.5$ and $E_0^* = 4G_0$ with $G_0$ being the instantaneous elastic shear modulus. In this special case, the normalized relaxation function is given by $\psi(t) = G(t)/G_0$, where $G(t)$ is the time-dependent shear modulus. In the general case, $\psi(t)$ can be expressed, for example, in terms of the shear relaxation modulus and the compression relaxation modulus (see, in particular, [7], Chapter 7).

It is interesting to observe that the rebound indentation displacement does not depend on the indenter shape, which, together with the fact that in the case of viscoelastic layer it does not depend on the layer thickness, means that the rebound indentation represents a robust indicator of the viscoelastic response of a tested time-dependent material. In other words, this property of being insensitive to the indenter shape and the sample size factors is found to be crucial for the reliability of indentation-based testing for biological materials [17].

Returning to the question of justification of the MDR in the viscoelastic case considered in the previous section, we would like to emphasize that the MDR applied in conjunction with Ting's algorithm [16] yields an explicit solution to the unilateral viscoelastic Hertz-type contact problem provided the time-dependent contact area is a singly-connected region with the radius being an arbitrary function of time. Meanwhile, the solution algorithm is not so important in view of the uniqueness of the solution. Indeed, let the histories $q(x,t)$, $a(t)$, and $w(t)$ solve the 1D viscoelastic contact problem (14)–(18), then separating the time interval $[0^-, t)$ into subintervals as $0^- < t_1 < t_2 < \ldots < t_n < t$, where $t_i$ are the roots of the equation $a(t_i) = a(t)$, we can represent the functions $q(x,t)$ and $u_z(x,t)$, $x \in (-a(t), a(t))$, in the form provided by Ting's algorithm, and thereby (due to the established equivalence) obtaining the unique solution to the original axisymmetric viscoelastic contact problem.

## 9 Conclusion

In the present paper, the MDR formalism has been extended for the case of Hertz-type viscoelastic contact. The application of the MDR to a unilateral contact problem for an arbitrary axisymmetric frictionless indenter and a linearly viscoelastic half-space under the assumption of circular contact area reduces the original contact problem to the corresponding unilateral 1D contact problem for the viscoelastic Winkler foundation and the equivalent rigid punch. By solving the equivalent 1D contact problem, one directly obtains both the relation between the contact force $F(t)$ and the indenter displacement $w(t)$ as well as the relation between the indenter displacement $w(t)$ and the contact radius $a(t)$, while the evaluation of the contact pressure density $p(r,t)$ requires application of the inverse MDR transformation mapping to the 1D viscoelastic foundation contact reaction $q(x,t)$.

**Acknowledgements** The authors are grateful to the DFG (German Science Foundation – Deutsche Forschungsgemeinschaft) for financial support.